\documentclass[12pt,preprint]{aastex}

\slugcomment{Not to appear in Nonlearned J., 45.}

\shorttitle{Absolute properties of the binary system BB Pegasi}
\shortauthors{B. Kalomeni et al.}

\received{2007 January 20}
\begin{document}

\title{Absolute properties of the binary system BB Pegasi}

\author{B. Kalomeni\altaffilmark{1}, K. Yakut\altaffilmark{2,3}, V. Keskin\altaffilmark{2}, \"O. L.De\u{g}irmenci\altaffilmark{2}, B. Ula\c{s}\altaffilmark{4,5}, and O. K\"ose\altaffilmark{2}}

\altaffiltext{1}{Department of Physics, \.Izmir Institute of Technology, Turkey }
\altaffiltext{2}{Department of Astronomy and Space Sciences, University of Ege, Turkey}
\altaffiltext{3}{Institute of Astronomy, K.U.Leuven, Belgium}
\altaffiltext{4}{Department of Physics, Onsekiz Mart University of \c{C}anakkale, Turkey}
\altaffiltext{5}{Department of Astrophysics, Astronomy and Mechanics, University of Athens, Greece}

\begin{abstract}
We present a ground based photometry of the low-temperature contact
binary BB~Peg. We collected all times of mid-eclipses available in
literature and combined them with those obtained in this study.
Analyses of the data indicate a period increase of $3.0(1)\times
10^{-8}$\,days/yr. This period increase of BB Peg can be interpreted
in terms of the mass transfer $2.4\times10^{-8}$ M$_{\odot}$
yr$^{-1}$ from the less massive to the more massive component. The
physical parameters have been determined as $M_\mathrm{c}$ = 1.42
M$_\odot$, $M_\mathrm{h}$ = 0.53 M$_\odot$, $R_\mathrm{c}$ = 1.29
R$_\odot$, $R_\mathrm{h}$ = 0.83 R$_\odot$, $L_\mathrm{c}$ = 1.86
L$_\odot$, and $L_\mathrm{h}$ = 0.94 L$_\odot$\ through simultaneous
solution of light and of the radial velocity curves.
 The orbital parameters of the
third body, that orbits the contact system in an eccentric orbit,
were obtained from the period variation analysis. The system is
compared to the similar binaries in the Hertzsprung-Russell and
Mass-Radius diagram.
\end{abstract}


\section{Introduction}

BB~Peg (HIP~110493, V = 11$^\mathrm{m}.6$ , F8V ) is a
low-temperature contact binary (LTCB) system which was
discovered as a variable star in 1931 by Hoffmeister.
Whitney (1959) refined the orbital period. Since then BB~Peg has
been the subject of several investigations. The system was observed
photoelectrically in 1978 by Cerruti-Sola \& Scaltriti (1980), Zhai
\& Zhang (1979), and Awadalla (1988). The times of minima of the system
have been published by numerous authors.
\par
Cerruti Sola, Milano \& Scaltriti (1981) analyzed the BV light curves of Cerruti-Sola \&
Scaltriti (1980) using the Wilson-Deviney (Wilson \& Devinney, 1971; Wilson, 1979; hereafter
WD) code. Giuricin, Mardirossian \& Mezzetti (1981) solved the same light curves using the Wood
(1972) model and obtained somewhat different results. Leung, Zhai \& Zhang (1985) used WD to
analyze the BV light curves obtained by Zhai \& Zhang (1979). Awadala (1988) observed UBV light
curves but did not perform a light curve analysis. The mass ratio was determined photometrically
for these light curve solutions. The first radial velocity study of the system
done by Hrivnak (1990) gives the mass ratio as 0.34(2). More recent radial velocity data
obtained by Lu \& Rucinski (1999) result in a mass ratio of 0.360(6). The photometric mass ratio (0.360$\mp$0.003) derived by Leung, Zhai \& Zhang (1985) agrees very well with the spectroscopic value, a result of the total/annular nature of the eclipses (see Terrell \& Wilson, 2005). Zola et al. (2003) published the physical parameters of the components. The orbital period variation was studied by Cerruti-Sola \& Scaltriti (1980) and Qian (2001).

\section{Observations}

The photometric observations of the system were obtained with the
0.4-m (T40), 0.35-m (T35) and 0.30-m telescopes (T30) at the Ege
University Observatory (EUO) and T\"UB\.ITAK National Observatory
(TUG) on 8 nights during the observing season between
August--December 2004 with T35 and 2 nights in 2006 with T40.
However, the system was observed at T30 and T40 for only three night
in order to obtain the minima times. The light curve of the system was
obtained from CCD photometry observations.
Light curves of BB Peg in Bessel V and R filters are shown in
Fig.~\ref{fig2}d and the data are given in Table\ref{tab1}. The comparison and check stars were
BD$+15^\circ4634$ and GSC 01682--01530, respectively.

\begin{table}
\caption{VR measurements of BB Peg (Fig.~\ref{fig2}d). The phases were calculated using the Eq. (1). 1 and 2 denote V, and R filters (F), respectively.} \label{tab1}
\begin{tabular}{llll}
\hline
\hline
HJD	&	Phase	&	$\Delta$m	&	F \\	
\hline
53301.3054	&	0.5115	&	1.2930	&	1	\\
53301.3067	&	0.5151	&	1.2840	&	1	\\
53301.3080	&	0.5187	&	1.2610	&	1	\\
53301.3093	&	0.5224	&	1.2500	&	1	\\
53301.3106	&	0.5260	&	1.2300	&	1	\\
53301.3119	&	0.5296	&	1.2050	&	1	\\
53301.3133	&	0.5333	&	1.1840	&	1	\\
53301.3146	&	0.5369	&	1.1690	&	1	\\
53301.3159	&	0.5405	&	1.1560	&	1	\\
\end{tabular}
\tablecomments{Table~\ref{tab1} is published in its entirety in the electronic edition of
the Astronomical Journal. A portion is shown here for guidance regarding
its form and content.}
\end{table}

\par
We obtained two minima times throughout these observations. They are
listed in Table~\ref{tab2} together with those published in existing
literature. Using these minima times we derived the linear
ephemeris:
\begin{equation}
HJD\,MinI = 24\,50657.4599(4)+0.3615015(1)\times E \label{bbpeq1}
\end{equation}
and used them in the reduction processes of the observed data.

\begin{table*}
\begin{center}
\scriptsize \caption{The primary (I) and the secondary (II) minima
times in HJD*(HJD - 2\,400\,000).}\label{tab2}
\begin{tabular}{lllllllll}
\hline
HJD$*$ & Min. & Ref.& HJD$*$ & Min. & Ref. & HJD$*$ & Min. & Ref.\\
\hline
26559.241   &   II  &   1   &   41181.397   &   I   &   5   &   50657.4575  &   I   &   16  \\
26582.014   &   II  &   2   &   41335.227   &   II  &   6   &   50671.3770  &   II  &   16  \\
26965.204   &   II  &   2   &   42405.259   &   II  &   7   &   50702.4698  &   II  &   17  \\
27393.223   &   II  &   2   &   42607.523   &   I   &   8   &   50739.7052  &   II  &   18  \\
30226.826   &   I   &   3   &   42748.310   &   II  &   8   &   50769.525   &   I   &   18  \\
30235.865   &   I   &   3   &   43729.4491  &   II  &   9   &   51078.4304  &   II  &   19  \\
30258.638   &   I   &   3   &   43730.3512  &   I   &   9   &   51471.3810  &   II  &   20  \\
30264.797   &   I   &   3   &   43754.3896  &   II  &   9   &   52131.8425  &   II  &   21  \\
30281.776   &   I   &   3   &   43754.3896  &   I   &   9   &   52201.2508  &   II  &   22  \\
30285.753   &   I   &   3   &   43757.4667  &   I   &   9   &   52201.4305  &   I   &   22  \\
30530.861   &   I   &   3   &   43764.3334  &   I   &   9   &   52203.2386  &   I   &   22  \\
30552.903   &   I   &   3   &   43806.08453 &   II  &   10  &   52203.4188  &   II  &   22  \\
30584.721   &   I   &   3   &   43806.98838 &   I   &   10  &   52207.3962  &   II  &   22  \\
30994.128   &   II  &   2   &   43813.13365 &   I   &   10  &   52513.4118  &   I   &   23  \\
31731.756   &   I   &   4   &   43814.03710 &   II  &   10  &   52838.402   &   I   &   24  \\
31783.455   &   I   &   4   &   43842.05373 &   I   &   10  &   52852.4956  &   I   &   25  \\
32433.631   &   II  &   4   &   43866.99893 &   I   &   10  &   53243.4607  &   II  &   26  \\
32433.801   &   I   &   4   &   44812.503   &   II  &   11  &   53284.3112  &   II  &   22  \\
32436.687   &   I   &   4   &   45208.3511  &   II  &   12  &   53285.3957  &   II  &   22  \\
32436.866   &   II  &   4   &   45208.5319  &   I   &   12  &   53353.3577  &   II  &   26  \\
32451.697   &   II  &   4   &   46024.2600  &   II  &   12  &   53984.3589  &   I   &   26  \\
32455.683   &   II  &   4   &   46026.2483  &   I   &   12  &   53984.3591  &   I   &   26  \\
32473.567   &   I   &   4   &   49243.4462  &   II  &   13  &   53984.5409  &   II  &   26  \\
32477.538   &   I   &   4   &   49244.3490  &   I   &   13  &   53984.5411  &   II  &   26  \\
32477.744   &   II  &   4   &   49273.2689  &   I   &   13  &   53986.5271  &   I   &   26  \\
32479.710   &   I   &   4   &   49275.2600  &   II  &   13  &   53986.5276  &   I   &   26  \\
34711.615   &   I   &   4   &   50001.3351  &   I   &   14  &   53992.4949  &   II  &   26  \\
35468.604   &   I   &   4   &   50026.2785  &   I   &   14  &   53992.4957  &   II  &   26  \\
36056.764   &   I   &   4   &   50359.4028  &   II  &   15  &           &       &       \\
\hline
\end{tabular}
\end{center}
{References for Table~\ref{tab2}. (1) Zessewitsch (1939), (2)
Tsessevich (1954), (3) Whitney (1943), (4) Whitney (1959), (5)
Diethelm (1973), (6) Locher (1973), (7) Diethelm (1976), (8)
Diethelm (1977), (9) Cerruti-Sola \& Scaltriti (1980), (10) Zhai \&
Zhang (1979), (11) Derman et al. (1982), (12) Awadalla (1988), (13)
M\"uyessero\u{g}lu et al. (1996), (14) Agerer \& H\"ubscher (1996),
(15) Agerer \& H\"ubscher (1998a), (16) Ogloza (1997), (17) Agerer
\& H\"ubscher (1998b), (18) Samolyk (1999), (19) Agerer et al.
(1999), (20) Agerer et al. (2001), (21) Nelson (2002), (22) Drozdz
\& Ogloza (2005a), (23) Demircan et al. (2003), (24) Bak\i\c{s} et
al. (2003), (25) H\"ubscher (2005), (26) Present study.}
\end{table*}

\section{Eclipse timings and period study}

The period variation study of the system was presented for the first
time by Cerruti-Sola \& Scaltriti (1980), resulting in the ephemeris Min I (HJD) $=2443764.3334(6)+0.3615021(2)E+2.3\times
10^{-11}E^2$. Qian (2001) presented it as Min I
(HJD)$=2430285.7618(6)+0.36150027(1)E+2.35(1)\times 10^{-11}E^2$.
\par

Recently, the existence of a third body was reported via spectroscopic
study by D'Angelo et al.(2006). We used the linear ephemeris given by
Qian (2001) to construct the binary's O-C diagram. It shows
almost a sine-like variation superposed on an upward parabola. A
sine-like variation in the O-C curve, where both the primary and the
secondary minima follows the same trend, suggests the light time
effect via the presence of a tertiary component.  Times of minima of
BB Peg yielded the following equation

 \begin{equation}
MinI = T_o +P_oE + \frac{1}{2} \frac{dP}{dE}E^2 + \frac {a_{12} \sin
i'}{c}  \left[ \frac {1-{e'}^2}{1+{e'} \cos v'} \sin \left( v' +
\omega' \right) +{e'} \sin \omega' \right]
 \label{bbpeq2}
\end{equation}
where $T_{\rm o}$ is the starting epoch for the primary minimum, $E$ is the integer
eclipse cycle number, $P_{\rm o}$ is the orbital period of the
eclipsing binary  $a_{12}$, $i'$ , $e'$, and $\omega'$
are the semi-major axis, inclination, eccentricity, and the
longitude of the periastron of eclipsing pair about the third body,
and $v'$ denotes the true anomaly of the position of the center of
mass. Time of periastron passage $T'$ and orbital period $P'$ are
the unknown parameters in Eq.(\ref{bbpeq2}).
\begin{table}
\caption{Orbital elements of the tertiary component in BB Peg. The
standard errors 1$\sigma$, in the last digit are given in
parentheses.} \label{tab3}
\begin{tabular}{lll}
\hline
Parameter            &Unit               & Value                       \\
\hline $T_o$         & [HJD]             & 2430285.7655(36)             \\
$P_o$                & [day]             & 0.3615006 (1)                 \\
$P'$                 & [year]            & 27.9(2.0)                        \\
$T'$                 & [HJD]             & 2438540 (793)                      \\
$e'$                 &                   & 0.56 (0.30)                          \\
$\omega'$            & [$^\circ$]        & 69 (18)                              \\
$a_{12} \sin i'$     & [AU]              & 0.96 (15)                               \\
$f(m)$               & [M$_{\odot}$]     & 0.0010(5)                                  \\
$m_{3;i'=10^\circ}$  & [M$_{\odot}$]     & 1.23                                        \\
$m_{3;i'=90^\circ}$  & [M$_{\odot}$]     & 0.16                                      \\
$Q$                  & [c/d]             & $1.5(2)\times 10^{-11}$              \\
\hline
\end{tabular}
\end{table}
\par
\begin{figure*}
\includegraphics{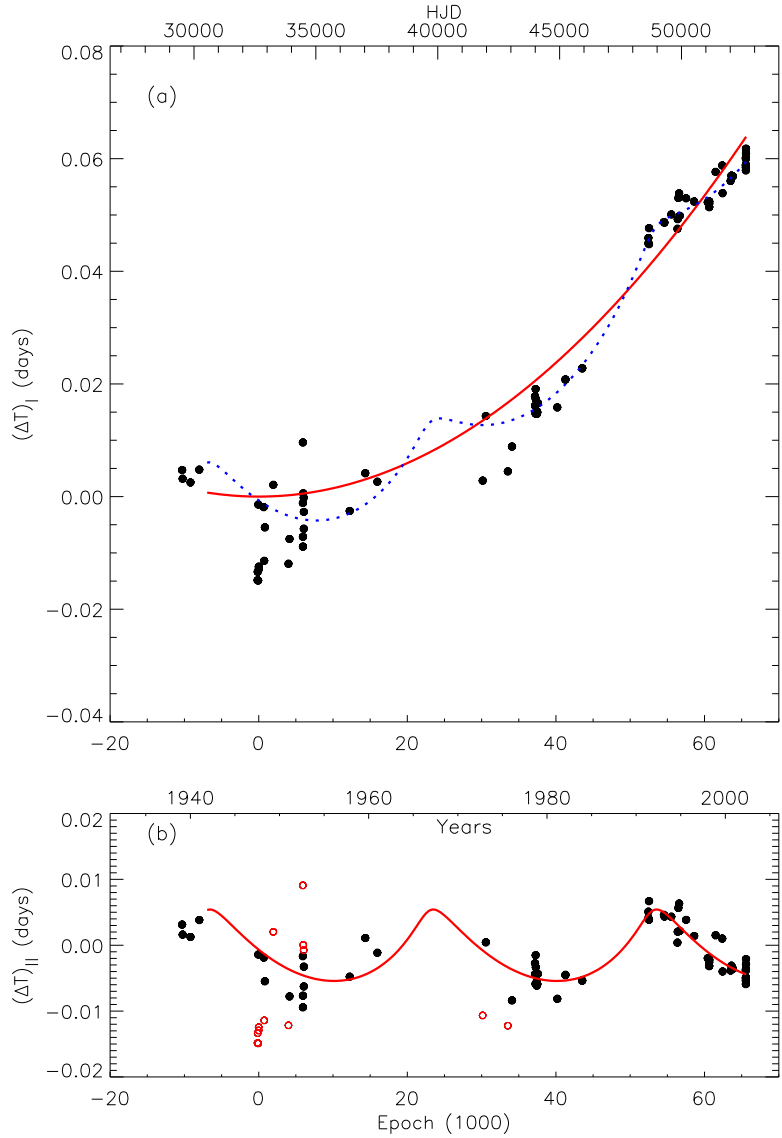} \caption{ (a) The $\triangle
T_{I}$ diagram of the times of mid-eclipses for BB Peg. The
continuous line is for the parabolic variation and the dashed line
is for a parabolic variation superimposed on the variation due to
the tertiary component, (b) the $\triangle  T_{II}$ residuals after
subtraction of parabolic change, shown in panel (a).}\label{fig1}
\end{figure*}

Light elements in Eq.(\ref{bbpeq2}) were determined using the
differential correction method. We used Eq.(2) along with the values
given in Table~\ref{tab2} and a weighted least squares solution to
derive the parameters shown in Table~\ref{tab3}. We assigned weight
10 to photoelectric (pe), 1 to photographic (pg) and 0 (pg) to a few
cases that shows high deviation from the expected normal position
(the open circles in Fig~\ref{fig1}b). The parameters given in
Table~\ref{tab3} were used for the $\triangle T_\mathrm{I}$ variation
study of the system which is plotted in Fig~\ref{fig1}a. The O-C
values in this figure were obtained with the linear elements
$T_\mathrm{o}$ and $P_\mathrm{o}$ given in Table~\ref{tab3}. The
line in Fig.~\ref{fig1}a shows the secular increase of the binary's
orbital period while the dashed line is for both the secular
increase and the light-time effect of the tertiary component. We
also present the contribution of the light-time effect, $\triangle
T_{\mathrm{II}}$, to total period variations of the system in
Fig~\ref{fig1}b. In the last section of this study we will discuss
the tertiary component in BB~Peg.

\section{Light curve analysis}
\begin{figure*}
{\includegraphics[height=13cm]{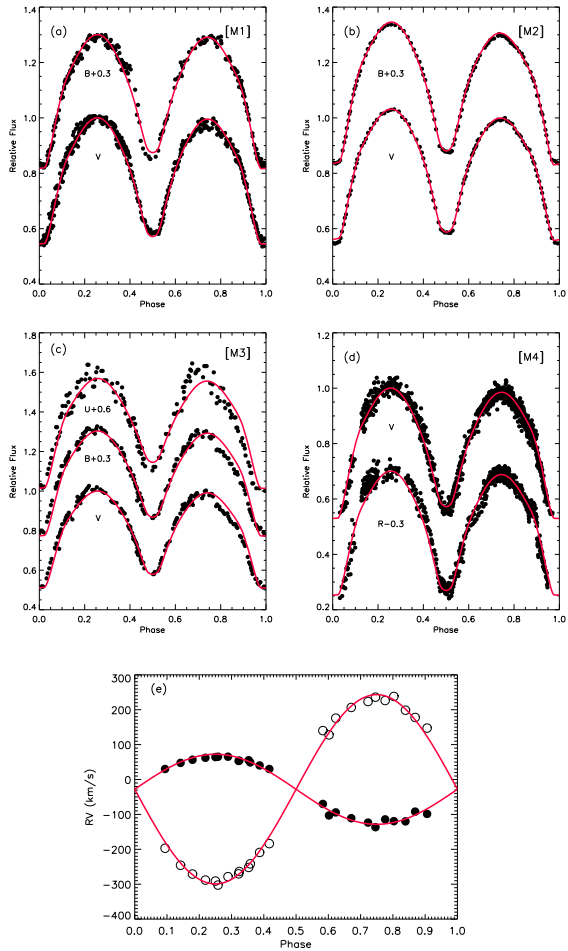}} \caption{ The observed and
computed light curves of BB~Peg. For the sake of comparison the
light curves in U, B, and R bands are shifted by a value of $+0.6$,
$+0.3$, and $-0.3$ in intensity. Radial velocities (e) from Lu \& Rucinski (1999)
and the computed curve through our simultaneous solution. See text for the details.}\label{fig2}
\end{figure*}

Previous light curves have been analyzed by
either old methods or with the assumption that photometric mass
ratio was known. All previously published light curves, as well as those of the
current study, have been analyzed simultaneously with the Lu and Rucinski (1999) radial
velocities using the latest version of the WD code (Wilson \& Devinney 1971; Wilson
1994). Mode 3 of the WD code has used throughout the
analysis. As seen in the Fig~\ref{fig2}, the light curves show asymmetries in the maxima.
Generally it is accepted that the stellar activity may cause to these asymmetries
in the light curves, we will discuss these asymmetries  in section 5.
Hence, the stellar spot parameters, were taken into consideration in our analysis.
The adopted values are: $T_\mathrm{1} = 6250$\,K
according to the $B-V$ color index, gravity darkening
coefficients, and albedos were chosen as $g_\mathrm{1}$ =
$g_\mathrm{2}$ = 0.32 (Lucy 1967) and $A_\mathrm{1}$ =
$A_\mathrm{2}$ = 0.5 (Rucinski 1969) and the logarithmic limb-darkening
coefficients ($x_\mathrm{1}$, $x_\mathrm{2}$) were obtained from
van Hamme (1993). Semi major axis of the relative orbit $a$, binary center-of-mass radial
velocity $V_\gamma$,  inclination $i$, temperature of the secondary
component $T_\mathrm{2}$, luminosities of the primary
component $L_\mathrm{1}$(U, B, V, R), the potential of the common
surface $\Omega$, and spot parameters (latitude, longitude, size and temperature factor) were adjustable parameters. The results are given in
Table~\ref{tab4}. Weights for the different sets of data were determined
by the scatter of the observations. In all analyses, the B, V and R filters were given 2 times
higher weight than the U filter to take their much better dispersion into account.
The computed light curves (solid lines) obtained
along with the parameters given in Table~\ref{tab4} were compared
with all observed light curves shown in Fig.~\ref{fig2}a, b, c, and
d. The synthetic light curves were created with the LC program.
\par
The obtained parameters for the light curves are given in
Table~\ref{tab4} . The results of the different light curve solution
models (M) have been denoted by different numbers. It has been
assigned M1 in Table~\ref{tab4} to two colors (B and V), light curves
solution obtained from Cerruti-Sola \& Scaltriti (1980), M2 to two
colors (B and V) light curves model of Zhai \& Zhang (1979) (the mean
values are taken from Leung et al. 1985), M3 to three colors (U, B,
and V) light curves that were obtained by Awadalla (1988) and,  M4 to
two color (R and V) light curves obtained in this study.
All the results appear to be compatible with each other.
Consistency of observations, using the
results given in Table~\ref{tab4}, with applied models are shown in
Fig.~\ref{fig2}a, b, c, and d.

\begin{table*}
\begin{center}
\caption{The photometric elements of BB Peg with their formal
1$\sigma$ errors. See text for details.}\footnotesize \label{tab4}
\begin{tabular}{llllll}
\hline
Parameter                            &  M~1          & M~2            &M~3          &M~4            \\
\hline
Geometric parameters:&&&\\
$i$ ${({^\circ})}$                   & 85.3(6)       & 87.9(1.4)      &84.6(9)    &85.0(5)            \\
$V_{\gamma}$                         & -27.8(1.7)    & -              &-28.0(2.0) &-28.1(2.2)         \\
$a$                                  & 2.665(30)     & -              &2.671(30)  &2.664(32)           \\
$\Omega _{1,2}$                      & 6.066(14)     & 6.045(6)       &6.005(20)  &6.056(13)           \\
$q$                                  & 2.752(27)     & -              &2.690(34)  &2.702(7)          \\
Filling factor (\%)                  & 35            & 38             &33         &34                 \\
Fractional radii of h. c.            &&&\\
${r}_{1~pole}$                       &0.2898(12)     & 0.2888(4)     &0.2922(15)   &0.2889(19)             \\
${r}_{1~side}$                       &0.3042(14)     & 0.3028(5)     &0.3068(19)   &0.3030(24)              \\
${r}_{1~back}$                       &0.3490(27)     & 0.3450(10)     &0.3522(34)  &0.3457(45)             \\
Fractional radii of c. c.            &&&\\
${r}_{2~pole}$                       &0.4541(11)     & 0.4499(4)     &0.4529(14)   &0.4507(16)               \\
${r}_{2~side}$                       &0.4894(15)     & 0.4839(5)     &0.4881(19)   &0.4849(22)               \\
${r}_{2~back}$                       &0.5209(20)     & 0.5145(7)     &0.5200(25)   &0.5157(30)               \\
Radiative parameters:&&&\\
$T_1$$^*$ (K)                        & 6250          & 6250           &6250         &6250            \\
$T_2$ (K)                            & 5905(45)      & 5945(40)       &5760(45)     &5955(30)         \\
Albedo$^*$ ($A_1=A_2$)               & 0.5           & 0.5            &0.5          &0.5              \\
Gravity brightening$^*$ ($g_1=g_2$)  &0.32           &0.32            &0.32         &0.32             \\
Luminosity ratio:$\frac{L_1}{L_1 +L_2}$(\%) &&&\\
$U$                                  & -             & -              &45           &-                    \\
$B$                                  & 37            & 36             &41           &-                     \\
$V$                                  & 36            & 34             &38           &34                     \\
$R$                                  & -             & -              &-            &32                     \\
Spot parameters: &&&\\
Colatitude                           & 1.24(7)       & 1.52(3)        &1.20(15)     &1.05(16)              \\
Longitude                            & 4.18(27)      & 4.36(6)        &4.26(40)     &4.78(29)                 \\
Spot radius                          & 0.18(2)       & 0.34(2)        &0.24(3)      &0.25(2)                   \\
Spot temperature                     & 0.92(2)       & 0.89(1)        &0.93(2)      &0.92(2)                  \\
\hline $^*$~Fixed
\end{tabular}
\end{center}
\end{table*}

Keeping in mind the possibility of a tertiary component orbiting
a third-body orbiting the binary system we assume the
3$^{\mathrm{rd}}$ body's ($l_3$) parameter as a free parameter
through the light curve solution. However, we couldn't find  meaningful values for the $l_3$
parameter throughout the solutions. Likewise, D'Angelo et al. (2006) showed that the light contribution
of the third body is tiny ($l_3/l_{1+2}=0.009$).

\section{Results and Discussion}
\begin{figure*}
\includegraphics{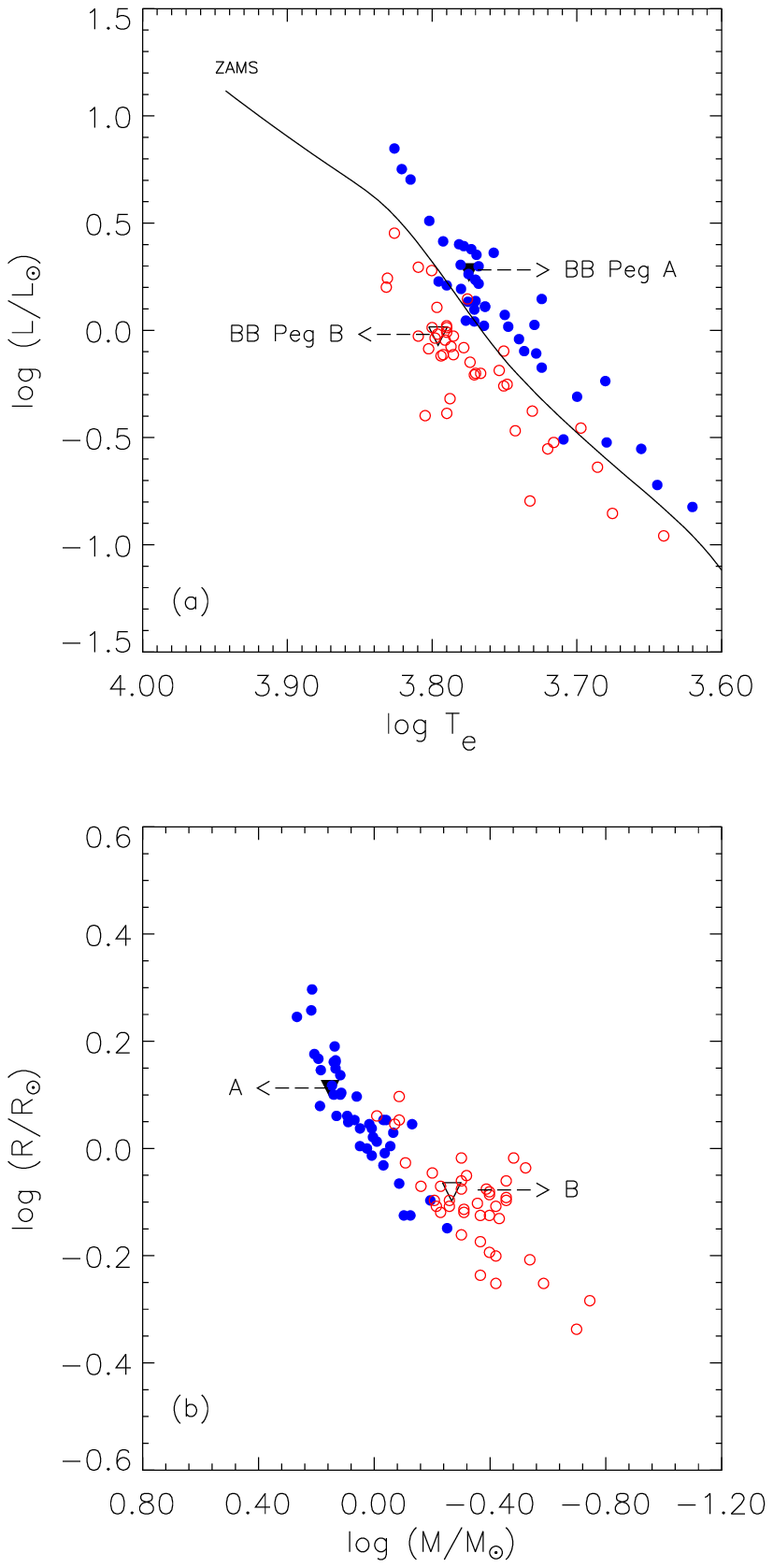}
\caption{The H-R and M-R diagrams showing BB Peg. The filled circles
show the primary component of W-type LTCBs, and the open circles are
for the secondaries. The ZAMS line is taken from Pols et al.
(1995).}\label{fig3}
\end{figure*}

All available light curves in literature, have been solved using the recent
WD code and the results are presented in Table~\ref{tab4}. The
solutions yielded very close results to each other. During the process effective
temperature and absolute magnitude of the Sun were taken as 5780~K
and 4.75 mag, respectively. In Fig.~\ref{fig3} the components
parameters are shown on the H-R and M-R diagrams. We show them along
with the LTCB systems (Yakut \& Eggleton 2005) whose physical
parameters are well-known. The results obtained from analyzing BB~Peg
(Table~\ref{tab5}) seem to be in good agreement with the
well-known LTCBs. Location of the less massive component in the
system indicates the system is overluminous and oversized, like the
other W-subtype secondary stars. Companion stars appear to be below
the ZAMS and the massive component is situated near the TAMS. If
interstellar absorption is not taken into account then through
the parameters given and using the values given in Table~\ref{tab5} the
distance of the system can be found as 361(25) pc. This is consistent with the HIPPARCOS value (ESA 1997).
The system's distance is derived from the Rucinski \& Duerbeck
(1997) in period-color-luminosity relation 389 pc,
 which is close to the one obtained in this study.
\begin{table*}
\caption{Absolute parameters of BB Peg. The standard errors
1$\sigma$ in the last digit are given in parentheses.}\label{tab5}
\begin{tabular}{lll}
\hline
Parameter (Unit)                                  & Hot component            & Cool component     \\
\hline
Mass ($\mathrm{M_{\odot}}$)                       & $0.53\,(2)$        & $1.42\,(4)$      \\
Radius ($\mathrm{R_{\odot}}$)                     & $0.83\,(2)$        & $1.29\,(2)$      \\
Effective temperature (K)                         & $6250$             & $5950\,(30)$      \\
Luminosity ($\mathrm{L_{\odot}}$)                 & $0.94\,(6)$        & $1.86\,(8)$       \\
Surface gravity (cgs)                             & $4.33$             & $4.37$            \\
Absolute bolometric magnitude (mag)               & 4.82$^{-0.08}_{+0.09}$& $4.08^{-0.13}_{+0.08}$ \\
Absolute visual magnitude (mag)                   & 4.98               & 4.26              \\
Distance (pc)                                    &\,\,\,\,\,\,\,\,\,\,\,\,\,\,\,\,$361^{-25}_{+20}$ &      \\
\hline
\end{tabular}
\end{table*}

Many contact binaries show an asymmetry in which one maximum is
higher than the other (the O'Connell effect), these asymmetries are
usually attributed to spots, which we interpret here in a very
general sense: they might be due to large cool star spots, to hot
regions such as faculae, to gas streams and their impact on the
companion star, or to some inhomogeneities not yet understood (Yakut
\& Eggleton). While the asymmetry is apparent in the shape of the
light curve of some systems (e.g. YZ Phe, Samec \& Terrell 1995),
), in others this asymmetry may not so prominent. (e.g, XY Leo,
Yakut et al. 2003). The asymmetry in the light curve of BB~Peg
is modeled with a cold spot on the secondary component (the cooler with higher mass and
radius) of the system. In the model of the light curve denoted by M2
the spot activity appears to be prominent with respect to the other
models. The results of the model are summarized in Table~\ref{tab4}. Besides, the
asymmetry in the light curve is well represented with the model (see
Fig.~\ref{fig2}).

Fig.~\ref{fig1} shows a parabolic variation. Therefore, we have
applied a parabolic fit, and assume that the mass transfer take
place between the components. The parabolic $(\triangle T_I)$ curve
shown in Fig.~\ref{fig1} indicates the existence of mass transfer in
the contact system, BB Peg. Upward parabolic variation suggesting a
mass transfer from the less massive component (the hotter component
in the case of BB Peg) to the more massive one. Eq.~(\ref{bbpeq2})
yields a period increase at a rate of $\frac{dP}{dt}=3.0(1)\times
10^{-8}$\,days yr$^{-1}$. If the period increase is indeed caused by
conservative mass transfer, then one can estimate the mass transfer
between the components. Using the derived masses, we derive the
rate of mass transfer $2.4(4)\times10^{-8}$ M$_{\odot}$ yr$^{-1}$ from
the less massive to the more massive component as in the
conservative mass transfer approximation. However, conservative mass transfer is just an optimistic
assumption. The non-conservative case is very important in close
binary evolution (see for details, Yakut \& Eggleton and
references therein). Analysis of the data, obtained over approximately 25 years, using the
WD program indicate a period increase of $2.9(1)\times
10^{-8}$\,days/yr, which is close to the one obtained with O-C
analysis.
\par
Fig.~\ref{fig1}b shows the variation of $\triangle T_{II}$ when the
observations extracted from the parabolic variation. The $\triangle
T_{II}$ variations show a sine-like variation, which implies the
existence of a tertiary component orbiting BB~Peg on an eccentric
orbit. Spectroscopic study of the system shows the existence of an M
-type dwarf star about the binary (D'Angelo et al. 2006). Using this
information with sine-line variation of the residuals of (O-C) we
solved the system under the assumption of existence of a third body and obtained
the values given in Table~\ref{tab3}. The results of the (O-C)
analysis show that the third component has a highly eccentric orbit
($e=0.56$) with about a 30-year period. Indeed, (O-C) residuals may
indicate that the source of this variation could be due to magnetic activity.
The orbit of third body, obtained in this study
compared to the data of Pribula \& Rucinski (2006) appeared to be much more
eccentric.
\par
On the other hand, using the values given in Table~\ref{tab3} and
Table~\ref{tab5} one may predict the mass of the tertiary
component. By assigning 0.96 AU to $a_{12}$sin$i$ and 29.7 yr to period
one can give the mass function as $0.0010 \, M_{\odot}$. For orbital
inclinations (i$_3$) of 90, 80, 50, 30, 10 the masses of the third
body ($m_3$)  are estimated to be 0.161, 0.164, 0.214, 0.341, 1.229
M$_{\odot}$, respectively. D'Angelo et al. (2006) found a
temperature of 3900~K for the tertiary component and a luminosity ratio
($\beta = \frac{l_3}{l_1+l_2}$) of 0.009. Following this
information with the deduced luminosities given in this study one
may give the radius of third body as 0.33 R$_{\odot}$. $M \simeq
0.978 R$ relationship is deduced using the ten well-known M-type
dwarf stars given in the study of Mercedes \& Ribas (2005), then
the tertiary body's mass of 0.32 M$_{\odot}$ is found. Taking into
consideration that value of mass, the orbital inclination of the
third body can be found as 35$^{\circ}$. Useful observations of BB
Peg throughout the next decade will help to determine the
accurate orbital parameters of the third body from the O--C diagram.

\acknowledgements We are grateful to C. Aerts, who has helped to
improve the final version of the paper with her comments and
suggestions. We thank Robert Smith for his suggestions, which improved the language of the manuscript.
We are very grateful to an anonymous referee for his/her comments and helpful constructive suggestions which
helped us to improve the paper. This study was supported by Ege University Research
Fund and T\"UB\.ITAK National Observatory. KY acknowledge support by the Research Council of the University of
Leuven under a DB fellowship.


\begin{thebibliography}{}
\bibitem[]{533} Agerer, F., Dahm, M., H\"ubscher, J.: 1999, IBVS~4712, 1

\bibitem[]{535} Agerer, F., Dahm, M., H\"ubscher, J.: 2001, IBVS~5017, 1

\bibitem[]{537} Agerer, F., H\"ubscher, J.: 1996, IBVS~4382, 1

\bibitem[]{539} Agerer, F., H\"uebscher, J.: 1998a, IBVS~4562, 1

\bibitem[]{541} Agerer, F., H\"ubscher, J.: 1998b, IBVS~4606, 1

\bibitem[]{543} Awadalla, N.~S.: 1988, Ap\&SS~140, 137

\bibitem[]{545} Bak{\i}\c{s}, V., Bak{\i}\c{s}, H., Erdem, A., \c{C}i\c{c}ek, C., Demircan, O., Budding, E.:
2003, IBVS~5464, 1

\bibitem[]{548} Cerruti-Sola, M., Milano, L., Scaltriti, F.: 1981, A\&A~101, 273

\bibitem[]{550} Cerruti-Sola, M., Scaltriti, F.: 1980, A\&AS~40, 85

\bibitem[]{552} D'Angelo, C., van Kerkwijk, M.~H., Rucinski, S.~M.: 2006, AJ~132, 650

\bibitem[]{554} Demircan, O., Erdem, A., \"Ozdemir, S., \c{C}i\c{c}ek, C., Bulut, \.I., Soydugan, F., Soydugan, E., Bak{\i}\c{s}, V., Kaba\c{s}, A., Bulut, A., T\"uys\"uz, M., Zejda, M., Budding, E.: 2003, IBVS~5364, 1

\bibitem[]{556} Derman, E., Y{\i}lmaz, N., Engin, S., Aslan, Z., Ayd{\i}n, C., T\"ufekcio\u{g}lu,
Z.: 1982, IBVS~2159, 1

\bibitem[]{559} Diethelm, R.:  1973, Rocznik Astron. Obser. Krakowskiego
44, 102

\bibitem[]{562} Diethelm, R.:  1976, Rocznik Astron. Obser. Krakowskiego 47,
96

\bibitem[]{565} Diethelm, R.:  1977, Rocznik Astron. Obser. Krakowskiego 48, 100

\bibitem[]{567} Drozdz, M., Ogloza, W.: 2005a, IBVS~5623, 1

\bibitem[]{569}ESA: 1997, The Hipparcos and Tycho Catalogues (ESA SP-1200).
ESA, Noordwijk

\bibitem[]{572} Giuricin, G., Mardirossian, F., Mezzetti, M.: 1981, AN~302, 285

\bibitem[]{574} Hoffmeister, C.: 1931, AN~242, 129

\bibitem[]{576} Hrivnak, B.~J.: 1990, BAAS~22, 1291

\bibitem[]{578} H\"ubscher, J.: 2005, IBVS~5643, 1

\bibitem[]{580} Leung, K.-C., Zhai, D., Zhang, Y.: 1985, AJ~90, 515

\bibitem[]{582} Locher, K.: 1973, SAC~44, 102

\bibitem[]{584} Lu, W., Rucinski, S.~M.: 1999, AJ~118, 515

\bibitem[]{586} Lucy, L. B. 1967, ZA~65, 89.

\bibitem[]{588}Mercedes, M-L, Ribas, I.: 2005, ApJ~631, 1120

\bibitem[]{590} M\"uyessero\u{g}lu, Z., G\"urol, B., Selam, S.~O.: 1996,
IBVS~4380, 1

\bibitem[]{593} Nelson, R.~H.: 2002, IBVS~5224, 1

\bibitem[]{595} Ogloza, W.: 1997, IBVS~4534, 1

\bibitem[]{597} Pols, O.~R., Tout, C. A., Eggleton, P.~P. Han, Z.: 1995, MNRAS~274, 964

\bibitem[]{599} Pribulla, T., Rucinski, S.~M.: 2006, AJ~131, 2986

\bibitem[]{601} Qian, S.: 2001, MNRAS~328, 635

\bibitem[]{603} Rucinski, S.~M.: 1969, AcA~19, 245

\bibitem[]{605} Rucinski, S.~M., Duerbeck, H.W.: 1997, PASP~109, 134

\bibitem[]{607} Samec, R.~G., Terrell, D.: 1995, PASP~107, 427

\bibitem[]{609} Samolyk, G.: 1999, AAVSO, 5

\bibitem[]{611} Terrell, D.,  Wilson, R.~E.: 2005, Ap\&SS~296, 221

\bibitem[]{613} Tsessevich, V.~P.: 1954, Odessa Izv. 4, 2, 271

\bibitem[]{615}van Hamme, W.: 1993, AJ~106, 2096

\bibitem[]{617}Whitney, B.~S.: 1943, AJ~50, 131

\bibitem[]{619} Whitney, B.~S.:1959, AJ~64, 258

\bibitem[]{621} Wilson, R.~E., Devinney, E. J.: 1971, ApJ~166, 605.

\bibitem[]{623} Wilson, R.~E.: 1994, PASP~106, 921.

\bibitem[]{625} Yakut, K., \.Ibano\u{g}lu, C., Kalomeni, B., De\u{g}irmenci, \"O.~L.: 2003, A\&A~401, 1095

\bibitem[]{627} Yakut, K., Eggleton, P.~P.: 2005, ApJ~629, 1055

\bibitem[]{629} Zessewitsch, W.: 1939, Quoted in Priceton Contribution
No. 19

\bibitem[]{632} Zhai, D. S., Zhang, Y.~X.: 1979, Xexue Tongbao 21, 895

\bibitem[]{634} Zola, S., Kreiner, J.~M., Zakrzewski, B., Kjurkchieva, D.~P., Marchev, D. V., Baran, A., Rucinski, S.~M., Ogloza, W., Siwak, M., Koziel, D., Drozdz, M., Pokrzywka, B.: 2005,
AcA~55, 389

\end{thebibliography}
\end{document}